\documentclass[reprint,twocolumn,superscriptaddress,preprintnumbers,amsmath,amssymb,aps,prd,floatfix]{revtex4-2}

\usepackage[utf8]{inputenc}

\usepackage{mathtools}
\usepackage{amsfonts}
\usepackage{mathrsfs}
\usepackage{bm}
\usepackage[normalem]{ulem}
\usepackage{bbold}
\usepackage{braket}
\usepackage{slashed}
\usepackage{dsfont}
\usepackage{float}
\usepackage{dcolumn}
\usepackage{amssymb,amsmath}

\usepackage{graphicx}
\usepackage{color}
\usepackage{array}

\usepackage{placeins}
\usepackage{booktabs}

\usepackage{xspace}
\usepackage{siunitx}
\DeclareSIUnit{\tops}{TOPS}
\DeclareSIUnit{\ops}{OPS}
\usepackage{hyperref}
\usepackage[nameinlink]{cleveref}
\usepackage{bookmark}

\usepackage{algorithm}
\usepackage{algpseudocode}
\DeclareMathOperator*{\argmin}{arg\,min}
\usepackage{enumitem}

\setkeys{Gin}{width=0.48\textwidth}

\usepackage{booktabs}
\usepackage{multirow}

\newcolumntype{C}{>{$}c<{$}}
\AtBeginDocument{
	\heavyrulewidth=.08em
	\lightrulewidth=.05em
	\cmidrulewidth=.03em
	\belowrulesep=.65ex
	\belowbottomsep=0pt
	\aboverulesep=.4ex
	\abovetopsep=0pt
	\cmidrulesep=\doublerulesep
	\cmidrulekern=.5em
	\defaultaddspace=.5em
}

\graphicspath{{./figures/}}

\usepackage{xifthen}
\usepackage{xcolor}

\newcommand{\gettitle}{Speeding up Fermionic Lattice Calculations with Photonic Accelerated Inverters}

\hypersetup{
	colorlinks,
	linkcolor={blue!75!black},
	citecolor={blue!75!black},
	urlcolor={blue!75!black}, 
	pdftitle={\gettitle},
	pdfauthor={},
	pdfkeywords={}
	{} 
	bookmarksopen=true,
	bookmarksopenlevel=2,
	bookmarksnumbered=true
}

\begin{document}
\title{\gettitle}

\author{Felipe Attanasio}
\affiliation{Institut f\"ur Theoretische Physik, Universit\"at Heidelberg, Philosophenweg 16, 69120 Heidelberg, Germany}

\author{Marc Bauer}
\affiliation{Institut f\"ur Theoretische Physik, Universit\"at Heidelberg, Philosophenweg 16, 69120 Heidelberg, Germany}

\author{Jelle Dijkstra}
\affiliation{Kirchhoff-Institut für Physik, Universität Heidelberg, Im Neuenheimer Feld 227, 69120 Heidelberg, Germany}

\author{Timoteo Lee}
\affiliation{Institut f\"ur Theoretische Physik, Universit\"at Heidelberg, Philosophenweg 16, 69120 Heidelberg, Germany}

\author{Jan M. Pawlowski}
\affiliation{Institut f\"ur Theoretische Physik, Universit\"at Heidelberg, Philosophenweg 16, 69120 Heidelberg, Germany}

\author{Wolfram Pernice}
\affiliation{Kirchhoff-Institut für Physik, Universität Heidelberg, Im Neuenheimer Feld 227, 69120 Heidelberg, Germany}

\begin{abstract}
    Lattice field theory (LFT) is the standard non-perturbative method to perform numerical calculations of quantum field theory. However, the typical bottleneck of fermionic lattice calculations is the inversion of the Dirac matrix. This inversion is solved by iterative methods, like the conjugate gradient algorithm, where matrix-vector multiplications (MVMs) are the main operation. Photonic integrated circuits excel in performing quick and energy-efficient MVMs, but at the same time, they are known to have low accuracy. This can be overcome by using mixed precision methods. In this paper, we explore the idea of using photonic technology to fulfil the demand for computational power of fermionic lattice calculations. These methods have the potential to reduce computation costs by one order of magnitude. Because of the hybrid nature of these methods, we call these 'photonic accelerated inverters (PAIs)'. 
\end{abstract}

\maketitle

\section{Introduction} 

Lattice field theory (LFT) is one of the standard non-perturbative approaches used to perform numerical calculations in quantum field theories. LFT enables the calculation of physical observables for various theories, allowing for the study of complex phenomena in strongly correlated systems such as phase transitions.  

Fermionic degrees of freedom make up the building blocks of many of the most intensely studied systems, including superconductors in condensed matter models, ultracold fermionic gases and quantum chromodynamics (QCD). 
Unfortunately, fermionic lattice calculations tend to be numerically expensive, as introducing fermions to calculations necessitates the use of algorithms which require many costly matrix operations, such as inversions. These inversions are usually facilitated via iterative approaches, like the Conjugate Gradient algorithm, by performing many matrix-vector multiplications (MVMs). They form the primary bottleneck for fermionic lattice calculations. As a result, large scale computational resources have become indispensable for such computations, in particular in lattice QCD, see e.g.~\cite{Ukawa:2015eka}. 

Photonic integrated circuits (PICs) are an emerging technology, offering a potential solution to meet the increasing need for computational power. This technology enables performing linear arithmetic operations using light instead of electrical currents with high throughput and low latency~\cite{Feldmann_2021,deLima:19}. PIC based hardware accelerators have the potential to deliver ultra-fast results~\cite{Shastri2021} and potentially ultra-low energy consumption~\cite{Wang2022,Chen2023}. 

PIC hardware is particularly suited to perform analog matrix-vector multiplications~\cite{Zhou2022}. This observation motivates this paper. The lattice community has always had an interest in building special purpose hardware, especially for LQCD calculations~\cite{Christ:1999ax,Chen:2000bu,Holmgren:2012wj,Hasenbusch:2003rs,Blum:2013mhx,Astrakhantsev:2020vhn,bates2020lattice} and PIC technology could be a good candidate for accelerating computational tasks. Furthermore, the application field of fermionic lattice calculations would provide an interesting alternative application beyond machine learning to the photonic computing community. 

However, one key limitation of photonic calculations is the low accuracy due to the limited dynamical range. Therefore, this paper explores using mixed precision methods which are known to work well on GPUs~\cite{Clark:2009wm}, in order to benefit from the speed and energy-efficiency provided by photonic accelerators. For this reason, the photonic calculations would not replace their digital counterparts, instead, they would complement them, hence, we call these photonic accelerated inverters (PAIs).  

The idea of using analog processors for iterative linear inverse solvers using mixed precision methods has been already proposed for general purposes~\cite{Zhu2023}. The goal of this paper is to use this idea to connect the lattice and photonic computing communities. For this reason, in \Cref{sec:lattice} we briefly review lattice field theory, and in \Cref{sec:fermions} we explain why fermionic lattice calculations are especially costly due to expensive matrix inversions. Then, in \Cref{sec:photonics} we review the potential and limitations of using photons for numerical calculations, and in \Cref{sec:simulation} we show how we simulate one of the main limitations of the PIC technology, which is the limited dynamical range. After the understanding gap between the communities is closed, we can introduce the CG algorithm in \Cref{sec:CG}, which is used to solve the matrix inversions, and mixed precision methods in \Cref{sec:mixed_precision} to benefit from the fast but inaccurate photonic calculations. Finally, in \Cref{sec:results} we show that these mixed precision algorithms can work well with the limited dynamical range of the photonic hardware for a $\varphi^4\text{-theory}$ with a Yukawa interaction, and we estimate the effective speedup compared to digital computations without photonic acceleration.

\section{Lattice Field Theory} 
\label{sec:lattice}

In this section, we briefly summarise the basics of lattice field theory. The idea of lattice field theory is to take advantage of the similarities between the path integral formulation of quantum field theory and statistical mechanics. This allows us to apply analytical and numerical methods, developed in statistical mechanics, to lattice field theory~\cite{Gattringer:2010zz}. 

For simplicity, we start with a scalar theory, namely, a $\varphi^4\text{-theory}$, where the Euclidean action is
\begin{equation}
    S_S[\varphi]=\int d^{d}\mathbf{x}\left[\frac{1}{2}\varphi(\mathbf{x})(-\Delta+m^{2})\varphi(\mathbf{x})+\frac{g_0}{4!}\varphi(\mathbf{x})^{4}\right] \,.
\end{equation}
This is the continuum action. However, digital computers can not handle continuous fields. Therefore, we need to use the discrete version of the action, which we call the lattice action. This is usually done over a hypercubic lattice in $d$ dimensions and spacing $a$ between the sites. The lattice action needs to yield the correct continuum limit ($a \to 0$) and conserve as many symmetries of the theory as possible. We can cast the lattice action of this theory into the following dimensionless form
\begin{equation}
    S_S = \sum_{n \in \Lambda} \left[ -2\kappa \sum_{\mu = 0}^d \varphi_n \varphi_{n+\hat \mu} + (1 - 2\lambda)\varphi_n^2  + \lambda \varphi_n^4  \right] \,,
    \label{eq:discrete_scalar_action}
\end{equation}
where $\varphi_n$ = $\varphi(x = na)$, $d$ is the number of dimensions, $\Lambda$ is the set of lattice points V = L$^{d}$ and $\hat \mu$ is a vector pointing to the next neighbour on the respective space-time direction. Using the path integral formulation, observables are given by
\begin{equation}
    \left\langle \mathcal{O} \right\rangle=\frac{1}{Z} \int \mathcal{D} [\varphi] e^{-S_S[\varphi]} \ \mathcal{O}[\varphi] \,, 
    \label{eq:path_integral}
\end{equation}
where
\begin{equation}
    Z =  \int \mathcal{D} [\varphi] e^{-S_S[\varphi]} \,,
\end{equation}
is the partition function and
\begin{equation}
    \mathcal{D} [\varphi] = \prod_{n \in \Lambda} d \varphi (n) \,,
\end{equation}
is the measure. The connection with statistical physics can be seen in \labelcref{eq:path_integral}, which takes the form of an expectation value with respect to a Boltzmann weight. Therefore, Markov chain Monte Carlo algorithms, like the Hybrid Monte Carlo algorithm in \Cref{fig:hmc}, can be used to generate $N$ field configurations $\varphi^{(i)}$, which follow the probability distribution
\begin{equation}
    p[\varphi^{(i)}] \propto e^{-S_S[\varphi^{(i)}]}\,.
    \label{eq:probability_distribution}
\end{equation}
Then, observables \labelcref{eq:path_integral} can be estimated as
\begin{equation}
    \left\langle \mathcal{O} \right\rangle \approx \frac{1}{N} \sum_{i = 1}^{N} \mathcal{O}[\varphi^{(i)}]\,.
    \label{eq:observables}
\end{equation}
%

\section{Fermionic Lattice Field Theory} \label{sec:fermions}

\begin{figure}[t]
    \centering
    \includegraphics{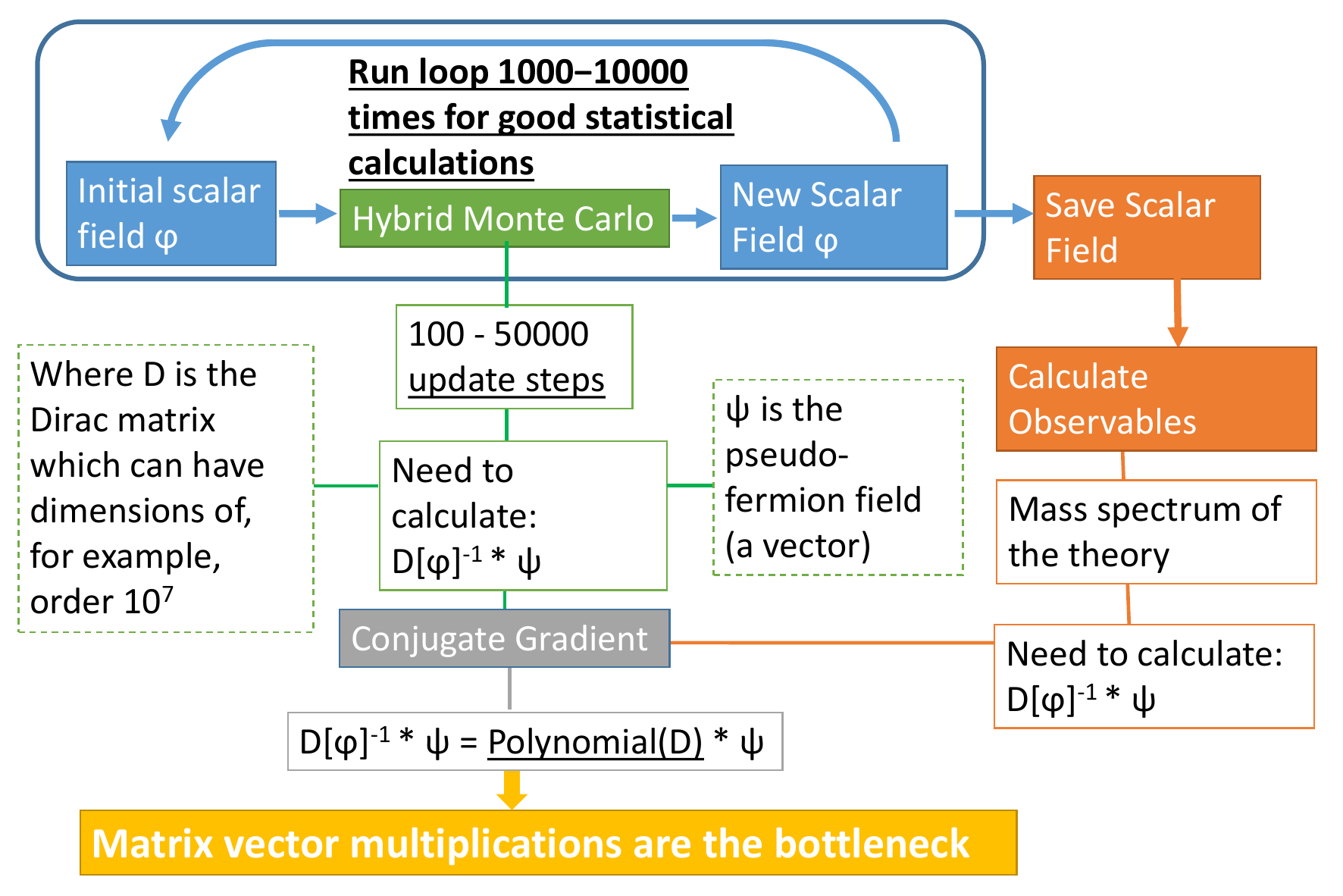}
    \caption{Schematic depiction of a typical lattice QFT simulation in fermionic theories.
    The main computational bottleneck lies in matrix-vector multiplications required to invert the Dirac matrix.}
    \label{fig:hmc}
\end{figure}
The purpose of this section is to summarise why we usually need to calculate the inversion of the Dirac matrix many times for fermionic lattice calculations. For simplicity, we use fermions coupled to the scalar field in \labelcref{eq:discrete_scalar_action} through a Yukawa coupling in 2 dimensions (1 space dimension and 1 time dimension) and with 2 mass-degenerate flavours. The Euclidean fermionic action with the Yukawa coupling reads
\begin{align}
    S_F = \sum_{f=1}^{2} \int dx^{\mu} & \Bigl( \overline{\psi}^{(f)} (x) (\gamma^\mu \partial_\mu + m) \psi^{(f)}(x) \nonumber\\ 
    &  + g \overline{\psi}^{(f)}(x) \varphi(x) \psi^{(f)} (x)\Bigr)\,,
    \label{eq:fermionic_action}
\end{align}
where $\mu \in \{0,1\}$, $\gamma$ are the Euclidean Dirac gamma matrices, $f$ is the index of the flavours, $g$ is the Yukawa coupling and $m$ is the degenerate fermion mass. The discretization of the fermionic action requires some care in order to avoid the doubling problem~\cite{Gattringer:2010zz,Rothe:1992nt}, which causes the lattice action to have the wrong continuum limit. One way of doing this is the Wilson formulation~\cite{Wilson:1974sk} with the action
\begin{equation}
    S_F = \sum_{f=1}^{2} \sum_{n,m} \bar{\psi}^{(f)}_m D_{mn} \psi^{(f)}_n \,,
    \label{eq:discrete_fermionic_action}
\end{equation}
where $\psi_n$ = $\psi(x = na)$ and the Dirac matrix is
\begin{align}
    D_{xy}=  & (m + 2 + g \varphi_y) \delta_{xy} \nonumber
    \\ &-\frac{1}{2}\sum_{\mu} \left(\Gamma_{\mu} \delta_{x+\hat\mu,y} + \Gamma_{-\mu} \delta_{x-\hat\mu,y} \right) \,,
    \label{eq:dirac_operator}
\end{align}
and $\Gamma_{ \pm \mu}=\left(1 \mp \gamma_\mu\right)$. The complications arise because the fermions $\psi$ obey Fermi statistics, which implies anti-symmetry under the interchange of quantum numbers. Thus all fermionic degrees anti-commute with each other. These anti-commuting numbers, like $\psi^{(f)}_n$ and $\bar{\psi}^{(f)}_m$ in \labelcref{eq:discrete_fermionic_action}, are the so-called Grassmann numbers. 
Numerically dealing with these anti-commuting numbers is challenging. To address the issue, observe that the fermionic part of the partition function 
\begin{equation}
    Z_F[\varphi] = \int \mathcal{D} [\psi, \bar{\psi}] \exp \left(-\sum_{f=1}^{2} \sum_{n,m} \bar{\psi}^{(f)}_m D_{mn}[\varphi] \psi^{(f)}_n \right),
\end{equation}
takes the form of the Gaussian integral over Grassmann numbers, which can be computed analytically. This leads to
\begin{equation}
    Z_F[\varphi] = \det D[\varphi] \det D^{\dagger}[\varphi] \,,
\end{equation}
where we have also used the property that $\gamma_5 D \gamma_5 = D^\dagger$, known as $\gamma_5$-hermiticity. This way, we integrated out the fermions $\psi$, and are left with the fermionic determinants. However, calculating these determinants directly is generally numerically expensive. Therefore, we can introduce an auxiliary bosonic field $\phi$ to write the fermionic determinants as a Gaussian integral with regular complex numbers 
\begin{equation}
    Z_F[\varphi] =  \int \mathcal{D} [\phi^{\dagger}, \phi] \exp \left(- \sum_{n,m} \phi^{\dagger}_m (D[\varphi] D^{\dagger}[\varphi])^{-1}_{mn} \phi_n \right) \,.
\end{equation}
This auxiliary field $\phi$ is called the pseudo-fermion field. Adding the scalar contribution in \labelcref{eq:discrete_scalar_action} to the action, the effective path integral is
\begin{equation}
    Z = \int \mathcal{D} [\varphi, \phi^{\dagger}, \phi] e^{-S_{\text{eff}}[\varphi, \phi^{\dagger}, \phi]} \,,
\end{equation}
where
\begin{equation}
    S_{\text{eff}}[\varphi, \phi^{\dagger}, \phi] = S_S[\varphi] +  \sum_{n,m} \phi^{\dagger}_m (D[\varphi] D^{\dagger}[\varphi])^{-1}_{mn} \phi_n\,,
    \label{eq:effective_action}
\end{equation}
is the effective lattice action. We can use the effective action to run a Markov Chain Monte Carlo algorithm, like Hybrid Monte Carlo~\cite{Gattringer:2010zz}, to generate fields configurations, which follow a probability distribution like in \labelcref{eq:probability_distribution}. This leads to algorithms like in \Cref{fig:hmc}, where we need to compute
\begin{equation}
    (D[\varphi] D^{\dagger}[\varphi])^{-1}_{mn} \phi_n
    \label{eq:inversion}
\end{equation}
for each field configuration to be generated by the algorithm. Note that one generally needs large numbers of field configurations to achieve good statistics. The inversion is usually computed using the conjugate gradient algorithm explained in \Cref{sec:CG}. This is why performing many matrix inversions is typically the bottleneck of fermionic lattice calculations.

\section{Photonic accelerators}
\label{sec:photonics}

In contrast to charged particles, photons can propagate through waveguides with barely any loss and have several exploitable degrees of freedom, such as wavelength and polarisation. This motivates using photons for numerical calculations. Photonic integrated circuits (PICs) based hardware accelerators particularly excel at linear arithmetic operations and many different architectures for photonic integrated matrix-vector multiplication processors have been proposed~\cite{Zhou2022,Xu2021,Feldmann_2021,Shen2017}. Therefore, there are different methods, which could benefit different fermionic lattice calculations. For example, meshed Mach–Zehnder interferometer (MZI) methods can be used to apply any unitary matrix to a vector~\cite{Reck:1994ujl,Clements:16}, which could be interesting for lattice QCD, because the Dirac operator is constructed with unitary matrices.

However, as a fundamental difference compared to their digital counterparts, the photonic accelerator chips represent numbers using analog signals, e.g. phases or light intensities, instead of bit patterns. Not only does this reduce the accuracy of the calculations, it also means that the calculations must use fixed point arithmetic where the full number range is divided into equally spaced steps. Furthermore, to interface with the existing digital computing infrastructure, the proposed analog photonic circuits typically employ digital-to-analog converters (DACs) at their inputs and analog-to-digital converters (ADCs) at their outputs. For this reason, the upper limit of the accuracy of the calculations is set by the limited range of values, which the DACs and ADCs can reach.

Such photonic integrated chips can then be treated as hardware accelerators for a general purpose processor, similar to a GPU that is used to offload graphics computations from the CPU. Many sources claim this technology could potentially outperform hardware accelerators based on digital electronics by potentially several orders of magnitude~\cite{Chen2023,8364605}. In this paper, we want to estimate the effective speedup which could potentially be reached using photonic processors and the required adaptions to the algorithms to cope with the limited accuracy.

\section{Simulation of the limited dynamical range}
\label{sec:simulation}

In the last section, we saw that the upper limit of the accuracy can be set by the limited dynamical range of the ADCs and the DACs. Therefore, simulating the behaviour of these components is a way to study ideal algorithms, which could overcome this limitation.  

The DACs can only reach discrete analog values. We assume that the DACs can reach integer values in the limited range [$-2^{\text{n}-1}$,~ 2$^{\text{n}-1} - 1$], where n is the number of bits of the DACs. To simulate this, we need to map the real numbers to integers. However, the extra steps required to perform this mapping can be numerically expensive. In particular, scalar-vector multiplications can be costly. To facilitate quicker computations, we make use of bit shifting, which is equivalent to multiplying the vector by scalar factors of $2^{k}$, where $k$ is a positive integer number. In our simulation, the scaling is done in three steps. They can be summarised as
\begin{enumerate}
    \item Find the maximum M of the absolute values of the vector.
    \item Find the shift $k = \text{n} - 1 + \text{ceil} (\log_2 \text{M})$ and perform a bit shift, where n is the number of bits of the DACs. This is equivalent to multiplying the factor $2^{k}$ to the vector.
    \item Round the numbers to integer numbers.
\end{enumerate}
This can be understood better through an example. If we have DACs with 6 bits and a vector \mbox{v~=~(-0.4168,~-0.1563,~-2.1362,~1.6403)}. The range of this DAC is \mbox{[$-2^5$,~$2^5-1$] = [$-$32,~31]} and the maximum absolute value of v is 2.1362. Then, the scaling factor is 8, leading to v~=~($-$3,~$-$1,~$-$17,~13), which are the mapped integer numbers in the range of the DACs. This is equally done for the matrix.

Additionally, ADCs can only read discrete values in a limited range [$-2^{\text{h}-1}$,~ 2$^{\text{h}-1} - 1$], where h is the number of bits of the ADCs. If the result of the operation is out of this range, then the maximum (minimum) value of the ADCs is read instead. This is called the saturation of the result. Therefore, after the matrix-vector multiplication is performed, the saturation is applied to the result by setting all values exceeding the maximum (minimum) value to the maximum (minimum) value. After the saturation is applied, the result has to be rescaled accordingly, to get the correct result with inaccuracies due to the limited dynamical range.

\section{Conjugate gradient} \label{sec:CG}
In \Cref{sec:fermions}, we discussed why matrix inversions are the bottleneck of fermionic lattice calculations. In this section, we explain one of the most used algorithms for these inversions, namely, the conjugate gradient (CG) algorithm. 

The CG algorithm is an iterative algorithm for solving $Ax = b$, where $A$ is a Hermitian and positive-definite matrix~\cite{Shewchuk1994AnIT}. In a nutshell, what the algorithm does can be summarised as finding the coefficients of the polynomial
\begin{equation}
    x = A^{-1}b = \sum_{i = 0}^{n - 1} c_i A^i b \,.
    \label{eq:CG}
\end{equation}
One can show that with infinite precision, there is always a finite polynomial, which exactly solves \labelcref{eq:CG} and the upper limit of $n$ is the dimension of $A$. However, because we can not get an exact result with finite precision, we usually have a stopping criteria, which depends on the chosen error tolerance. 

The inspiration for the derivation of this algorithm comes from the fact that the solution of $Ax = b$ is also the minimum of the quadratic function 
\begin{equation}
    f(x) = \frac{1}{2} x^T A x - b^T x\,,
    \label{eq:quadratic_function}
\end{equation}
which only has one minimum for Hermitian positive-definite matrices. In deep learning, one algorithm used to minimise the loss function is the stochastic gradient descent algorithm, which is ideal for neural networks because it works for all differentiable functions. However, having a specific function like in \labelcref{eq:quadratic_function}, it is more efficient to find an algorithm made specifically for finding its global minimum. One of these specific algorithms is the CG algorithm.  
\begin{figure}[t]
    \centering
    \includegraphics{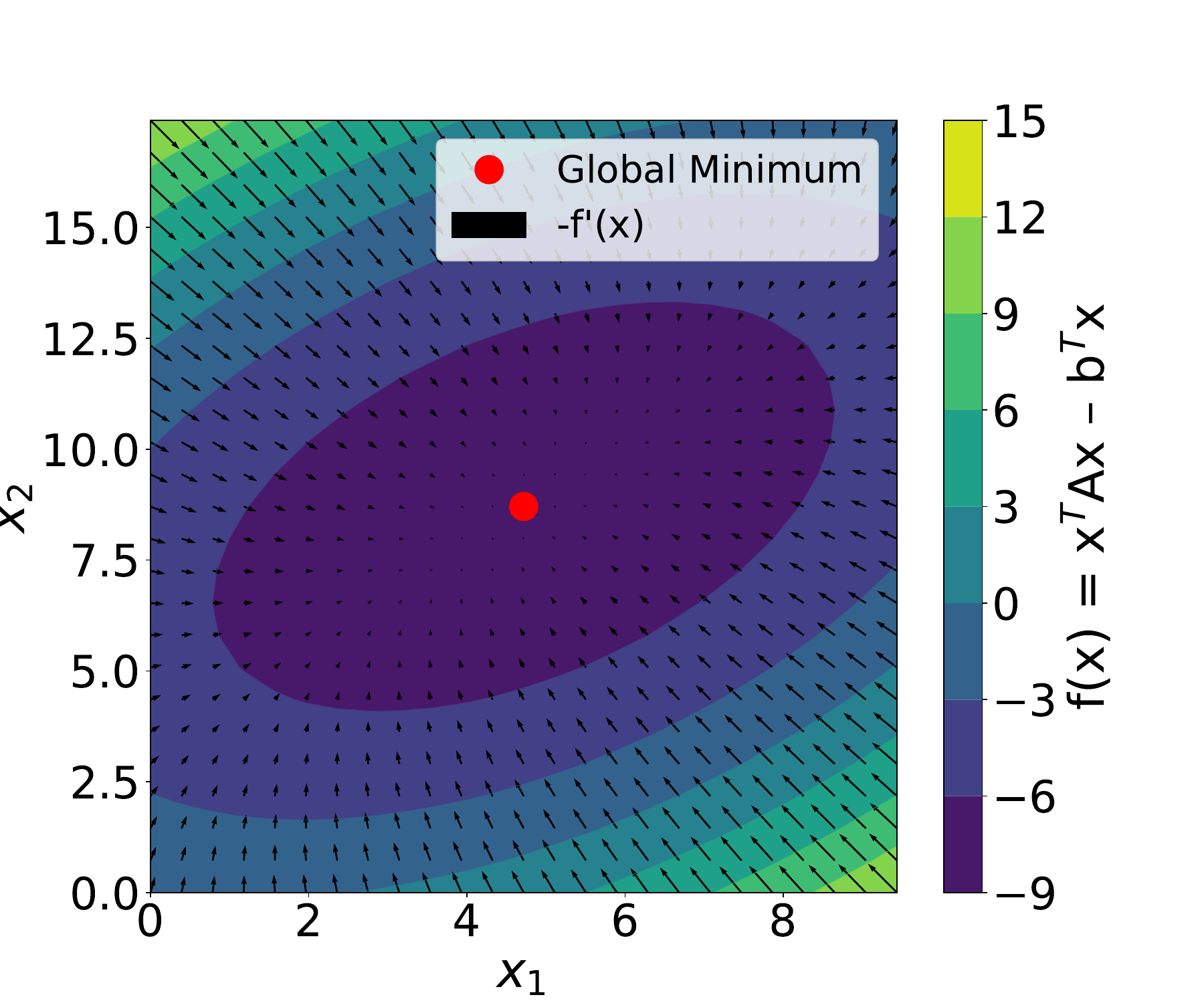}
    \caption{Visualisation of the residual in \labelcref{eq:residual} at different positions $x$ for a 2 by 2 matrix $A$. We can see that moving along these directions of steepest descent, we can move closer to the global minimum.}
    \label{fig:vectorfield}
\end{figure}

\begin{figure}[t]
    \centering    \includegraphics{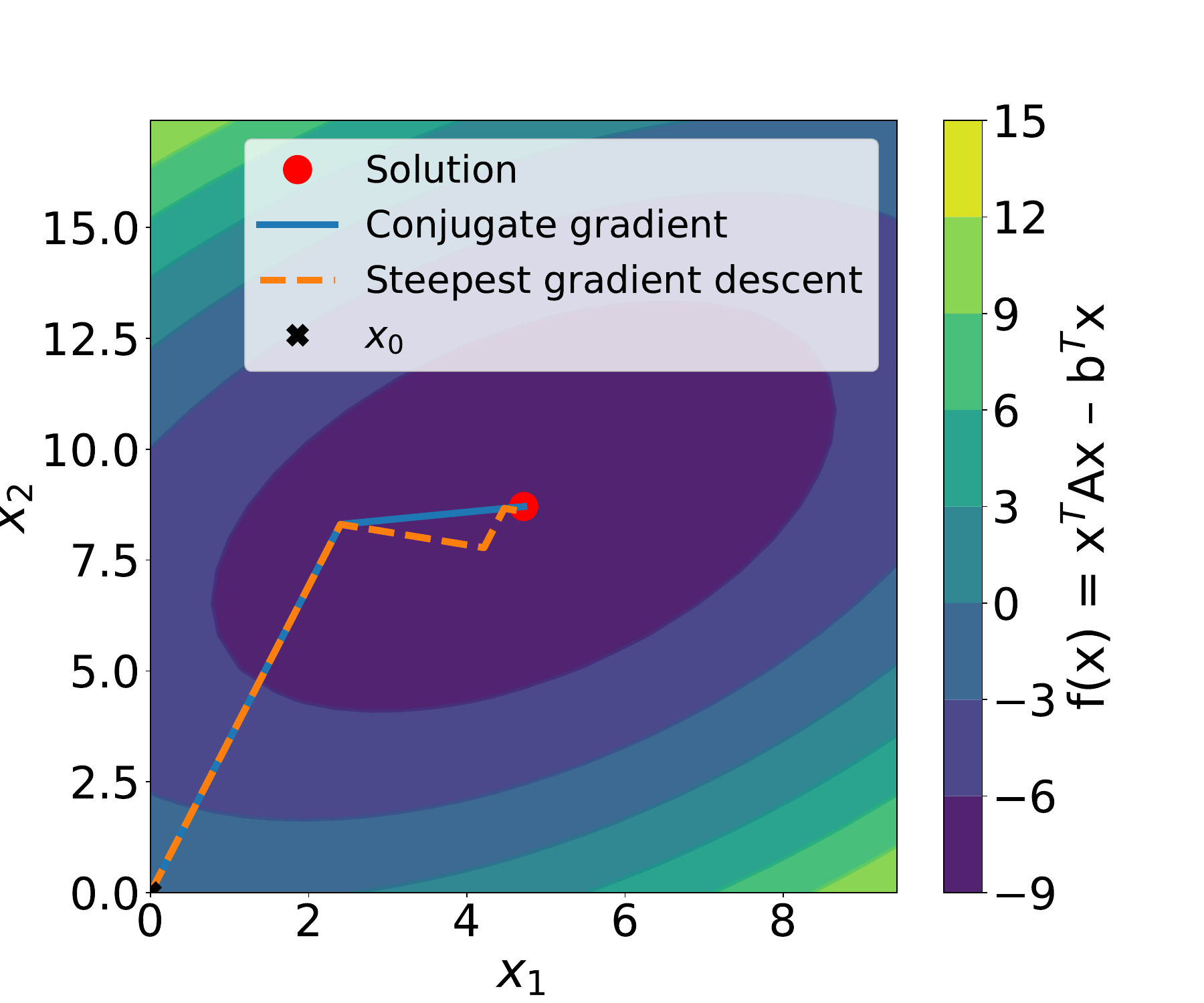}
    \caption{Comparison between the steepest gradient descent algorithm and the conjugate gradient algorithm. $x_0$ is the initial guess. We can see that the steepest gradient descent algorithm moves in this zigzag path and CG avoids this zigzag motion.}
    \label{fig:SGD_CG}
\end{figure}

The main idea of any gradient descent algorithm is the fact that $-f'(x)$ points to the direction in which $f(x)$ decreases the most quickly, as can be seen in \Cref{fig:vectorfield}. In other words, $-f'(x)$ is the direction of the steepest descent. For the quadratic function, this is equal to
\begin{equation}
    r_{(k)}:= r(x_{(k)}) := - f'(x_{(k)}) = b - Ax_{(k)} \,,
    \label{eq:residual}
\end{equation}
which is also called the residual $r$. This residual indicates how far we are from the correct solution. Because $r$ is the direction of the steepest descent, we can move closer to the solution with
\begin{equation}
    x_{(k+1)} = x_{(k)} + \alpha r_{(k)} \ .
    \label{eq:update_sgd}
\end{equation}
In stochastic gradient descent, $\alpha$ is the learning rate, and this is a hyperparameter of the algorithm. In the steepest gradient descent algorithm, the algorithm chooses $\alpha$ according to
\begin{equation}
    \alpha = \argmin_{\alpha} f(x_{(k)} + \alpha r_{(k)}) \ .
    \label{eq:optimal_alpha}
\end{equation}
One interesting property of choosing such $\alpha$ is
\begin{equation}
    \langle r_{(k + 1)} , r_{(k)} \rangle = r_{(k + 1)}^T r_{(k)} = 0, \ \forall k \in \mathbb{N} .
    \label{eq:orthogonal}
\end{equation}
This means that $r_{(k + 1)}$ is orthogonal to the previous residual $r_{(k)}$. By iterating this, we will eventually converge to the solution, which can be seen in \Cref{fig:SGD_CG} for a 2 by 2 positive-definite symmetric real matrix $A$. However, we can notice here a zigzag path. This is not efficient, because we move along the same direction multiple times and it would be ideal if we would move only once along each direction.

The conjugate gradient algorithm avoids this issue by introducing search directions $p$, which are not orthogonal like $r$ in \labelcref{eq:orthogonal}, but instead they are conjugate with respect to $A$
\begin{equation}
    \langle p_{(i)}, p_{(j)} \rangle_A = p_{(i)}^T A p_{(j)} = 0, \ \forall i,j \in \mathbb{N} \ . 
    \label{eq:conjugate_orthogonal}
\end{equation}
With such a set of search directions $p$, and if $\alpha$ is again calculated such that
\begin{equation}
    \alpha = \argmin_{\alpha} f(x_{(k)} + \alpha p_{(k)}) ,
    \label{eq:optimal_alpha_conjugate}
\end{equation}
we avoid moving along the same direction more than once, like can be seen in \Cref{fig:SGD_CG}. We find these conjugate search directions $p$ using the conjugate Gram-Schmidt method
\begin{align}
    p_{(k + 1)} = r_{(k + 1)} + \sum_{l= 0}^{k} \beta_{kl} p_{(l)} \,.
    \label{eq:gram_schmidt}
\end{align}
We can simplify the calculation of $r$ and $\beta$ in such a way that it becomes an efficient iterative algorithm and we get \Cref{alg:CG}. In summary, we find a set of search directions $p$, which are conjugate. We make sure this property is fulfilled by using the residuals $r$ and the conjugate Gram-Schmidt algorithm. $\alpha$ is calculated to minimise \labelcref{eq:optimal_alpha_conjugate}, and the values of $\alpha$ fulfil
\begin{equation}
    x = x_0 + \sum_{j = 0}^{n-1} \alpha_{j} p_j \,,
    \label{eq:summary_CG}
\end{equation}
where the upper limit of $n$ is the number of dimensions of the matrix $A$. This way, the zigzag path in the steepest gradient descent method is avoided and we get an algorithm, which can converge in fewer iterations.  

\begin{figure}
\begin{algorithm}[H]
\caption{The Conjugate Gradient algorithm}\label{alg:CG}
\begin{algorithmic}
\State $r = b - Ax$ \Comment{$x$ is the initial guess, usually set to 0}
\State $p = r$
\State k = 0
\State $R_0 = r^Tr$ \Comment{Initial residue square}
\State $R = R_0$
\While{($R > \epsilon R_0$) and (k $<$ iterMax)}\Comment{$\epsilon$ and iterMax}
    \State $\alpha = R/p^TAp$ \hspace{1.6cm} need to be chosen empirically
    \State $x \gets x + \alpha p$
    \State r $\gets$ r - $\alpha Ap$ 
    \State $R^{\prime} = R$
    \State $R = r^Tr$
    \State $\beta = R/R^{\prime}$ 
    \State $p \gets r + \beta p$
    \State k $\gets$ k + 1
\EndWhile
\State \textbf{return} x
\end{algorithmic}
\end{algorithm}
\end{figure}
\section{Mixed precision conjugate gradient}\label{sec:mixed_precision}

The conjugate gradient algorithm is guaranteed to converge with infinite precision~\cite{Shewchuk1994AnIT}. However, this is not the case for finite precision, especially, for the limited dynamical range of an analog hardware, as one can see in \Cref{fig:example2d_0}. In this section, we introduce mixed precision methods to achieve the desired accuracy.

The idea of the mixed precision conjugate gradient (MPCG) algorithm is to benefit from quick but inaccurate matrix-vector multiplications and ensure high accuracy in the inversion by performing high-precision corrections from time to time. How to perform this correction step is not trivial and hence, there are different mixed precision methods~\cite{Clark:2009wm}. We use in this paper the iterative refinement method, also called the defect-correction method, which works not only for CG but for any low-precision linear solver.

\subsection{Iterative refinement}
\label{sec.ItRefine}

\begin{figure}[t]
    \centering    \includegraphics{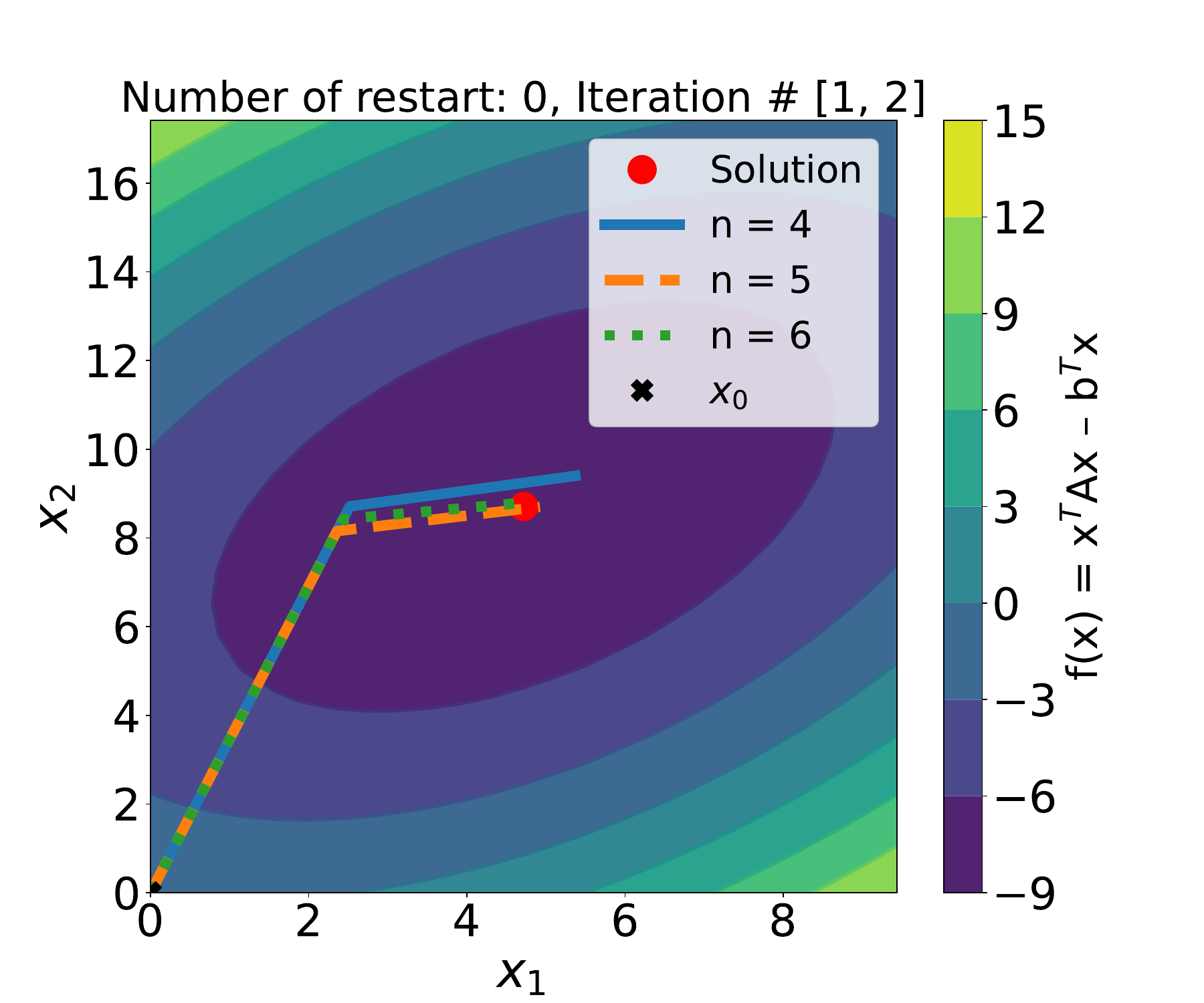}
    \caption{Visualisation of CG runs with different fixed point bit precision n: Because the initial guess is x = 0, all runs have the same first search direction $p$. The algorithm deviates due to the low accuracy. However, all the runs moved closer to the solution.}
    \label{fig:example2d_0}
\end{figure}

In \Cref{fig:example2d_0}, we can see that CG does not converge to the correct solution with low-precision calculations, but it does move closer to the result. The idea of the iterative refinement method is to benefit from this fact and move again closer to the result by resetting the calculation with a high-precision step~\cite{moler67}. The algorithm does this as follows
\begin{enumerate}[ref={Step~\arabic*}]
    \item Choose initial guess $x$. Usually set to $\vec{0}$  
    \item Calculate $r = b - Ax$ with high precision
    \item Solve $Ay = r$ with low precision CG with the accuracy defined by $\epsilon^{\text{low}}$, where the search direction $p$ needs to be scaled efficiently at each iteration. \label{enum:low_precision}
    \item Update $x$ $\gets$ $x + y$
    \item Iterate 2 to 4 until $Ax=b$ is solved with the accuracy defined by $\epsilon^{\text{final}}$
\end{enumerate}
Therefore, we reset the calculation by calculating the residual $r$ with high precision and then solving \mbox{$Ay=r(x)$}. This can be seen as using low-precision CG to find a new guess $x$ closer to the solution, resetting the calculation, and iterating this until the desired accuracy is reached. Note that if the inner loop is solved perfectly, the algorithm would converge in one update step because
\begin{align}
    &Ay = r(x) = b - Ax \nonumber\\
    &\Rightarrow A(x + y) = b.
\end{align}

\subsection{Condition number}
\label{sub:condition}

\begin{figure}[t]
    \centering
    \includegraphics{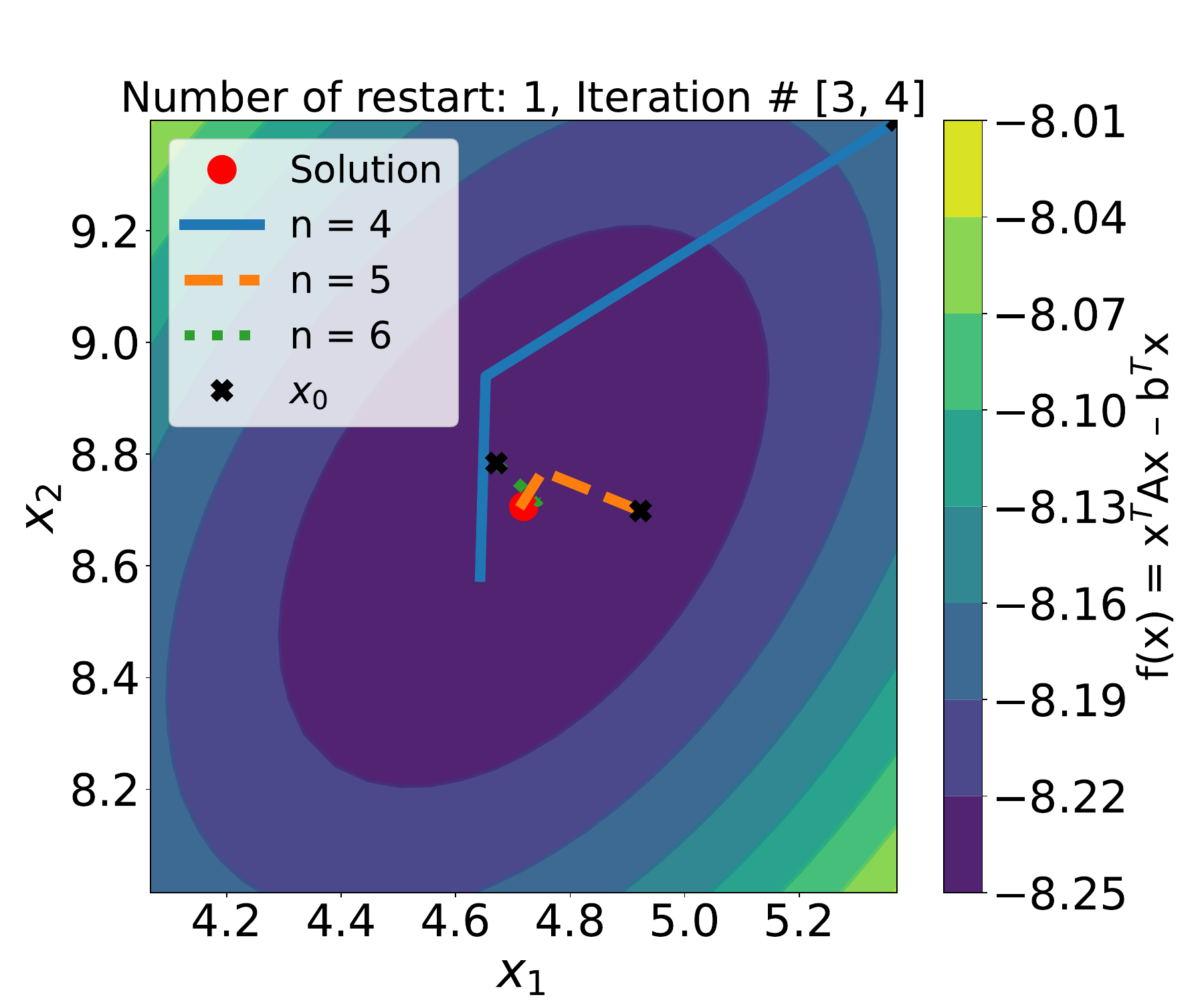}
    \caption{Continuing from the guesses $x$ in \Cref{fig:example2d_0}, we reset the algorithm. After two low-precision iterations, again, the algorithm comes closer to the solution. This is repeated until convergence.}
    \label{fig:example2d_1}
\end{figure}
The convergence rate depends on the eigenvalues of the matrices. For example, one function of the eigenvalues, which is known to affect the numerical stability of CG, is the condition number~\cite{Shewchuk1994AnIT}
\begin{equation}
    C(A) = \frac{EV_{\text{max}}(A)}{EV_{\text{min}}(A)} ,
    \label{eq:condition_number}
\end{equation}
where $EV_{\text{max}}(A)$ and $EV_{\text{min}}(A)$ are the maximal and minimal eigenvalues of $A$, and they are positive because $A$ is positive-definite. This will explain the trend in the number of iterations needed for convergence in the results in \Cref{sec:results}. 

We can optimise the behaviour of the eigenvalues by using preconditioners \cite{peardon2000accelerating}, which is essential for iterative linear solvers, and improve the convergence speed. Choosing optimal preconditioners will be crucial for optimising the photonic accelerated inverters and this needs to be investigated in the future. 

Furthermore, variations of the iterative refinement method have been proposed, which are guaranteed to not diverge and could further improve the stability of the mixed precision methods against bad condition numbers~\cite{wu2023stable}.

\subsection{Calculation time}
\label{sec:CalcTime}

To estimate the calculation time of the methods introduced here, we assume that only the matrix-vector multiplications (MVMs) have a significant contribution to the calculation time. Then, the calculation time for CG is
\begin{equation}
    t^{(s)} = N_d^{(s)} T_d ,
    \label{eq:time_standard}
\end{equation}
where $N$ is the number of iterations needed to converge to the desired accuracy, $T$ is the duration of one MVM in units [time/\#iterations], $s$ is the superscript for standard CG and $d$ is the subscript for double precision calculations. Analogously, the computation time of MPCG is
\begin{equation}
    t^{(mp)} = N_d^{(mp)} T_d + N_l^{(mp)} T_l ,
    \label{eq:time_mixed}
\end{equation}
where $mp$ is the superscript of MPCG and $l$ is the subscript for lower precision calculations. We can see that the limit of infinite hardware speed is equivalent to $T_l \to 0$. Therefore, the lower limit of the calculation time of MPCG is
\begin{equation}
    t_{\text{limit}}^{(mp)} = N_d^{(mp)} T_d < t^{(mp)}.
\end{equation}
This means that no matter how much quicker the photonic calculations are, the calculation time has a clear lower limit. It is important to note that this is the same for digital accelerators, like GPUs.

\subsection{Multilayered MPCG}
\label{sec:multilayered}

We can see that we can improve the lower limit of the calculation time by reducing the number of double precision correction steps $N_d$. The multilayered mixed precision conjugate gradient does this by using multiple layers of mixed precision calculations~\cite{Carson2018AcceleratingTS}. For example, using the photonic hardware as the lowest precision, single precision as the intermediate precision, and double precision as the highest precision. The idea is to solve $Ay=r$ in~\ref{enum:low_precision} by calculating a new residual $l(y) := r - Ay$ with intermediate-precision, solving $Az = l$ with low-precision with an accuracy defined by $\epsilon^{\text{low}}$, updating $y \gets y + z$, and iterating this until the accuracy set by $\epsilon^{\text{in}}$ is reached.

Using this multilayered algorithm, the calculation time is determined by 
\begin{equation}
    t^{(ml)} = N_d^{(ml)} T_d + N_i^{(ml)} T_i + N_l^{(ml)} T_l,
    \label{eq:time_multi}
\end{equation}
where $ml$ is the superscript for multilayered and $i$ is the subscript for intermediate precision. With this, the lower limit of the calculation time is
\begin{equation}
    t_{\text{limit}}^{(ml)} = N_d^{(ml)} T_d + N_i^{(ml)} T_i < t^{(ml)}.
    \label{eq:upper_limit2}
\end{equation}
where the $N_d^{(ml)}$ here is expected to be smaller than $N_d^{(m)}$ in \labelcref{eq:time_mixed}.

\subsection{Effective speedup factor}
\label{sec:speedup}

We want to calculate the effective speedup factor gained by using the multilayered MPCG with photonic acceleration compared to running MPCG without photonic acceleration. For the runs with MPCG, we use half precision as the lower precision, and for the run with the multilayered MPCG, we use single precision as the intermediate precision and the simulated photonic hardware as the lowest precision. To simplify the analysis, we define
\begin{equation}
    S_i = \frac{T_d}{T_i} \quad \text{and} \quad S_l = \frac{T_d}{T_l} \quad \text{and} \quad 
    I = \frac{N_l}{N_d^{(s)}} ,
    \label{eq:definitions}
\end{equation}
where $S$ is the hardware speedup factor respective to the calculation done in double precision and $I$ is the increase factor of the number of iterations relative to standard CG. Then, the effective speedup factor is
\begin{equation}
    f_{\text{eff}} = \frac{t^{(mp)}}{t^{(ml)}} = \frac{N_d^{(mp)} + N_d^{(s)} I^{(mp)} S_h}{N_d^{(ml)} + N_i^{(ml)} S_s + N_d^{(s)} I^{(ml)} S_p} ,
    \label{eq:effective_speed2}
\end{equation}
where $h$ is the subscript for half precision, $s$ is the subscript for single precision and $p$ is the subscript for the photonic accelerator.

\section{Results}
\label{sec:results}

\begin{figure}[t]
    \centering
    \includegraphics{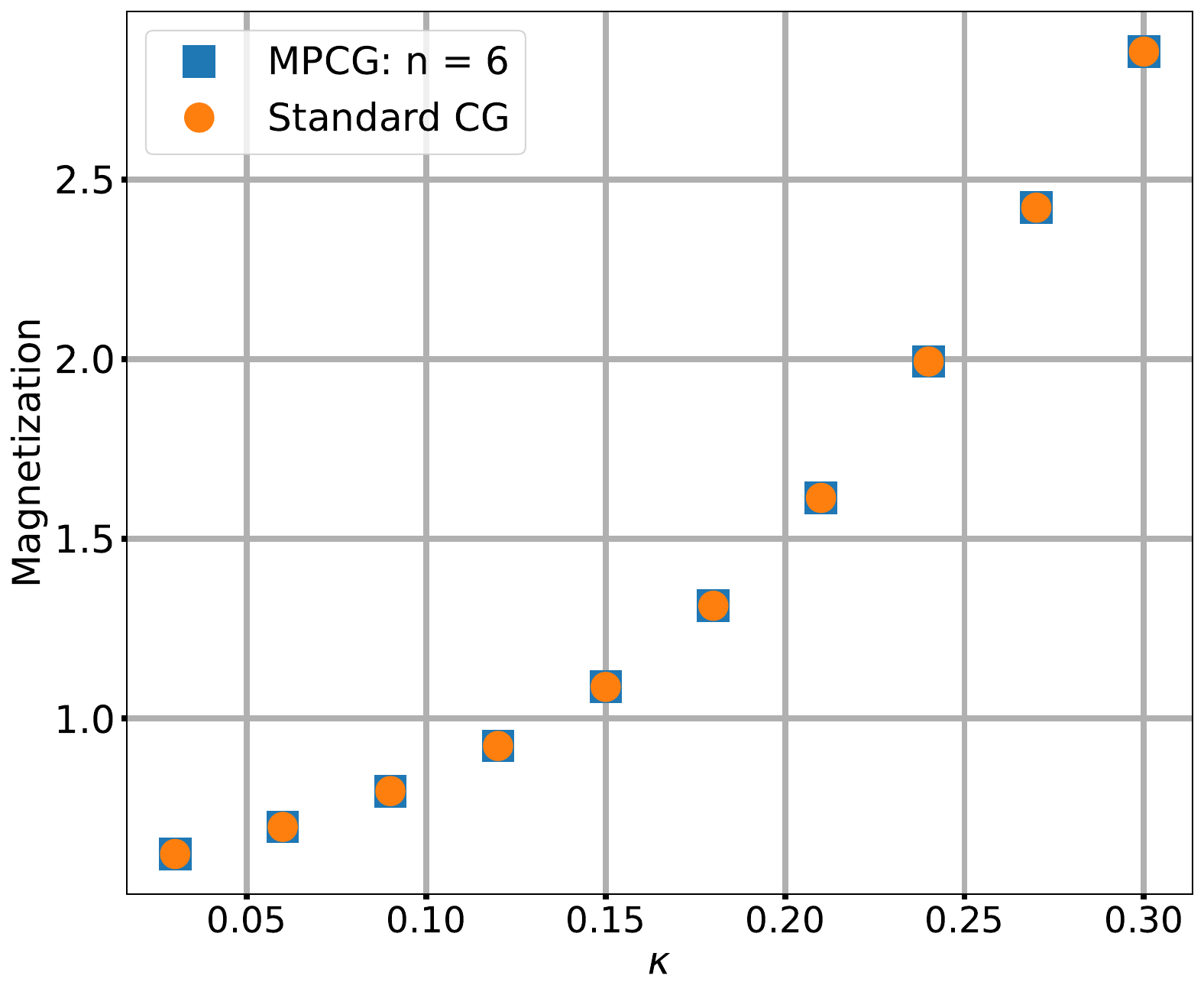}
    \caption{The simulation in \Cref{fig:hmc} was performed for a 64x64 lattice, $\lambda = 0.02$, $m = 1$, $g = 1$ and different $\kappa$ values. This was done with standard CG and with MPCG with DACs with 6 bit precision. The magnetisation in \labelcref{eq:magnetization} was calculated for each value of $\kappa$.}    
    \label{fig:magnetization}
\end{figure}
In \Cref{sec:fermions} we derived the effective action \labelcref{eq:effective_action}, with $S_S$ in \labelcref{eq:discrete_scalar_action} and $D$ in \labelcref{eq:dirac_operator}. Using this effective action for performing HMC, we get an algorithm which is described in \Cref{fig:hmc}, where we need to solve \labelcref{eq:inversion} many times. In this section, we show that the algorithms introduced here can work for the limited dynamical range of the hardware but at the cost of higher number of iterations. Then, we show the potential of using these photonic accelerated inverters to alleviate the typical bottleneck of fermionic lattice calculations.

To check that the mixed precision methods are working correctly, we can use the magnetisation
\begin{equation}
    M[\varphi] = \frac{1}{V} \sum_{n \in \Lambda} \varphi_n \, .
    \label{eq:magnetization}
\end{equation}
Running the simulation with the exact same conditions with MPCG and with standard CG yields coinciding magnetisation values, which indicates that all inversions were performed correctly. An example can be seen in \Cref{fig:magnetization}.

\subsection{Number of iterations}
\label{sec:NumberIts}

\begin{figure}[t]
    \centering
    \includegraphics{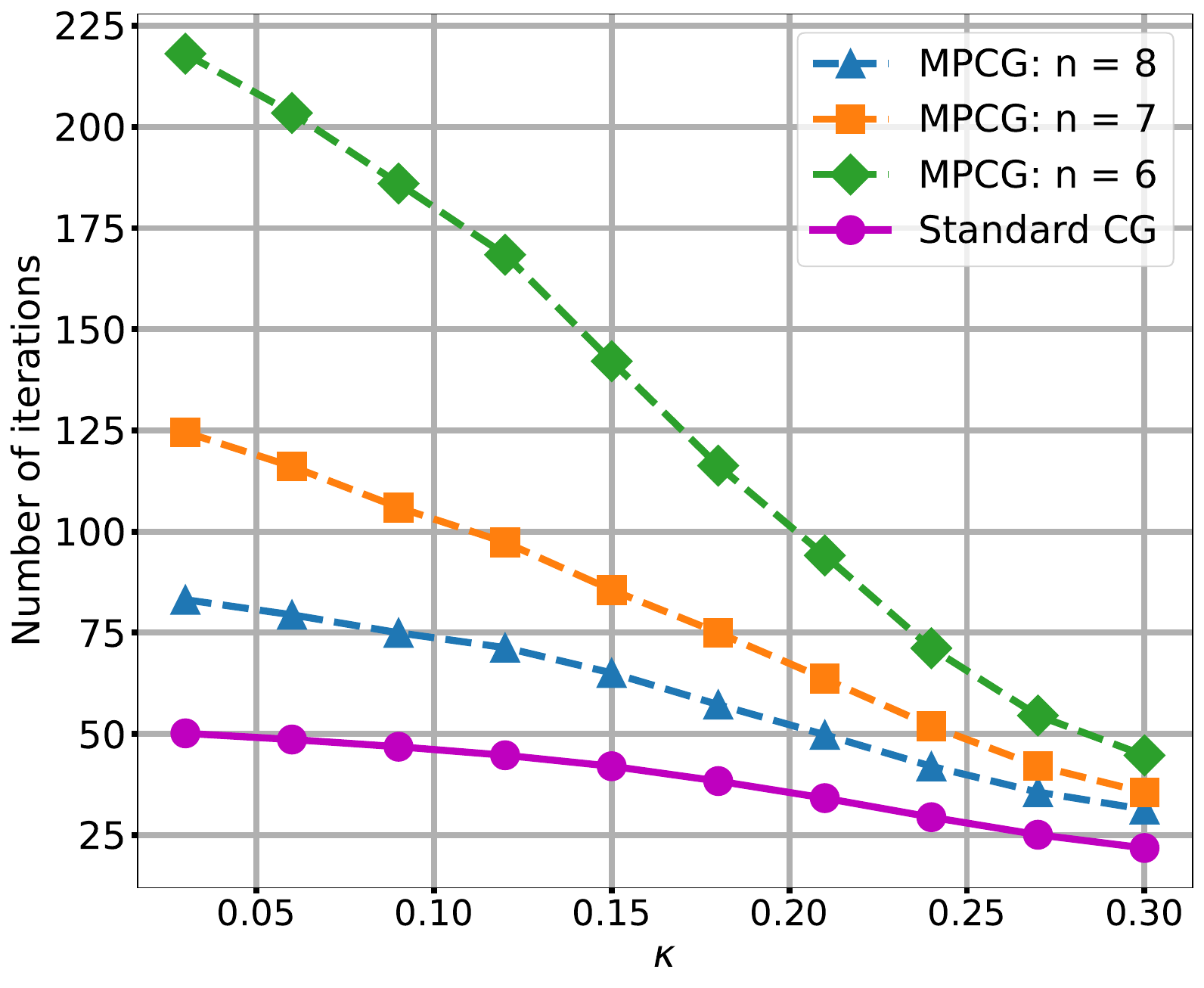}
    \caption{The average number of iterations needed in each inversion in \Cref{fig:magnetization} is plotted as a function of $\kappa$. n is the number of bits of the DACs.}    
    \label{fig:iterations}
\end{figure}
In \Cref{fig:iterations}, we plot the average number of iterations (sum of the number of low precision iterations and the high precision correction steps) needed for the MPCG algorithm to converge with DACs with different n fixed-point bit precision and with the stopping criteria set by $\epsilon^{\text{low}}=2^{-2(\text{n}-1)}$ and $\epsilon^{\text{final}}=10^{-12}$, and this is compared to standard CG with the same final stopping criteria. We performed in total 12126 inversions for each value of $\kappa$. The bit precision of the ADCs is set to 14, except for the run with the DACs with 8-bit precision, where the bit precision of the ADCs is set to 16. The errors have been omitted for visualisation purposes, and these are not relevant to our analysis and conclusion. 

As expected, the mixed precision algorithm needs more iterations to converge to the desired accuracy than standard CG. The number of iterations increases for smaller $\kappa$ values because $\kappa$ affects the condition number \labelcref{eq:condition_number} of the Dirac matrix, which affects the numerical stability of CG in general. We can see that this effect is stronger for the runs with lower accuracy.

This shows that mixed precision methods could potentially overcome the limited dynamical range of the photonic calculations and enable future collaboration between the photonic and the lattice communities.

\subsection{Effective Speedup of Multilayered MPCG} 
\label{sec:EffSpedup}

Now we want to estimate the possible effective speedup gained using the proposed photonic accelerated inverters. To show the potential of these photonic accelerated inverters, we use the NVIDIA A100 as an example. The NVIDIA A100 can perform calculations with single precision 8 times faster than with double precision and with half-precision 16 times faster~\cite{9361255}. Furthermore, many sources claim that the photonic computations will outperform their digital counterparts by many orders of magnitude~\cite{Chen2023,8364605}. For the sake of demonstration, we assume that this means that in the future, the photonic computation will be 1000 times faster than the half-precision calculations, hence, 16000 faster than the double precision calculations of the NVIDIA A100. This leads to $S_s$~=~8, $S_h$~=~16, $S_p$~=~16000. 

\begin{table}[t]
\begin{tabular}{|c|c|c|c|c|c|c|c|}
\hline 
$\kappa$ & $\langle N_d\rangle$ & $\langle N_i\rangle$ & $\sigma (N_i)$ & $\langle N_l\rangle$ & $\sigma (N_l)$ & $\langle I\rangle$ & $f_{\text{eff}}$ \\
\hline
0.03 & 2.00 & 6.41 & 1.07 & 116.45 & 50.26 & 2.32 & 3.24 \\
\hline
0.06 & 2.00 & 6.32 & 0.81 & 109.39 & 43.74 & 2.25 & 3.22 \\
\hline
0.09 & 2.00 & 6.24 & 0.70 & 102.18 & 38.79 & 2.18 & 3.20 \\
\hline
0.12 & 2.00 & 6.21 & 0.62 & 96.81 & 36.44 & 2.17 & 3.15 \\
\hline
0.15 & 2.00 & 6.11 & 0.41 & 86.40 & 27.96 & 2.06 & 3.11 \\
\hline
0.18 & 2.00 & 6.04 & 0.28 & 74.36 & 20.72 & 1.94 & 3.04 \\
\hline
0.21 & 2.00 & 6.01 & 0.11 & 63.66 & 12.40 & 1.87 & 2.95 \\
\hline
0.24 & 2.00 & 6.00 & 0.01 & 53.01 & 6.23 & 1.80 & 2.84 \\
\hline
0.27 & 2.00 & 6.00 & 0.00 & 43.93 & 3.22 & 1.75 & 2.75 \\
\hline
0.30 & 2.00 & 6.00 & 0.00 & 38.22 & 2.22 & 1.75 & 2.67 \\
\hline
\end{tabular}
\caption{The number of iterations needed by the multilayered MPCG with 8-bit precision. The superscript $ml$ has been omitted in this table.}
\label{table:1}
\end{table}
Using this, we want to estimate the effective speedup factor in \labelcref{eq:effective_speed2}. 
For the MPCG calculation time, we assume for simplicity $I^{(mp)} = 1$ and $N_d^{(mp)} = 6$, and we take the values of $N_d^{(s)}$ from \Cref{fig:iterations}. 
We use the multilayered MPCG to run the same simulation like in \Cref{fig:iterations} with 8 fixed-point bit precision and with single precision as the intermediate precision. 
The results with $\epsilon^{\text{low}} = 6\cdot 10^{-5}$ for the low precision calculation,  $\epsilon^{\text{in}} = 10^{-7}$ for the intermediate precision, and  $\epsilon ^{\text{final}}= 10^{-12}$ are shown in \Cref{table:1} and this leads to the estimated effective speedup factors shown there. 
We can see that the estimated effective speedup factors for our Yukawa theory in two dimensions range from 2 to 4 and we notice that the effective speedup factor improves as the matrix is harder to invert. 

\begin{figure}[t]
    \centering
    \includegraphics{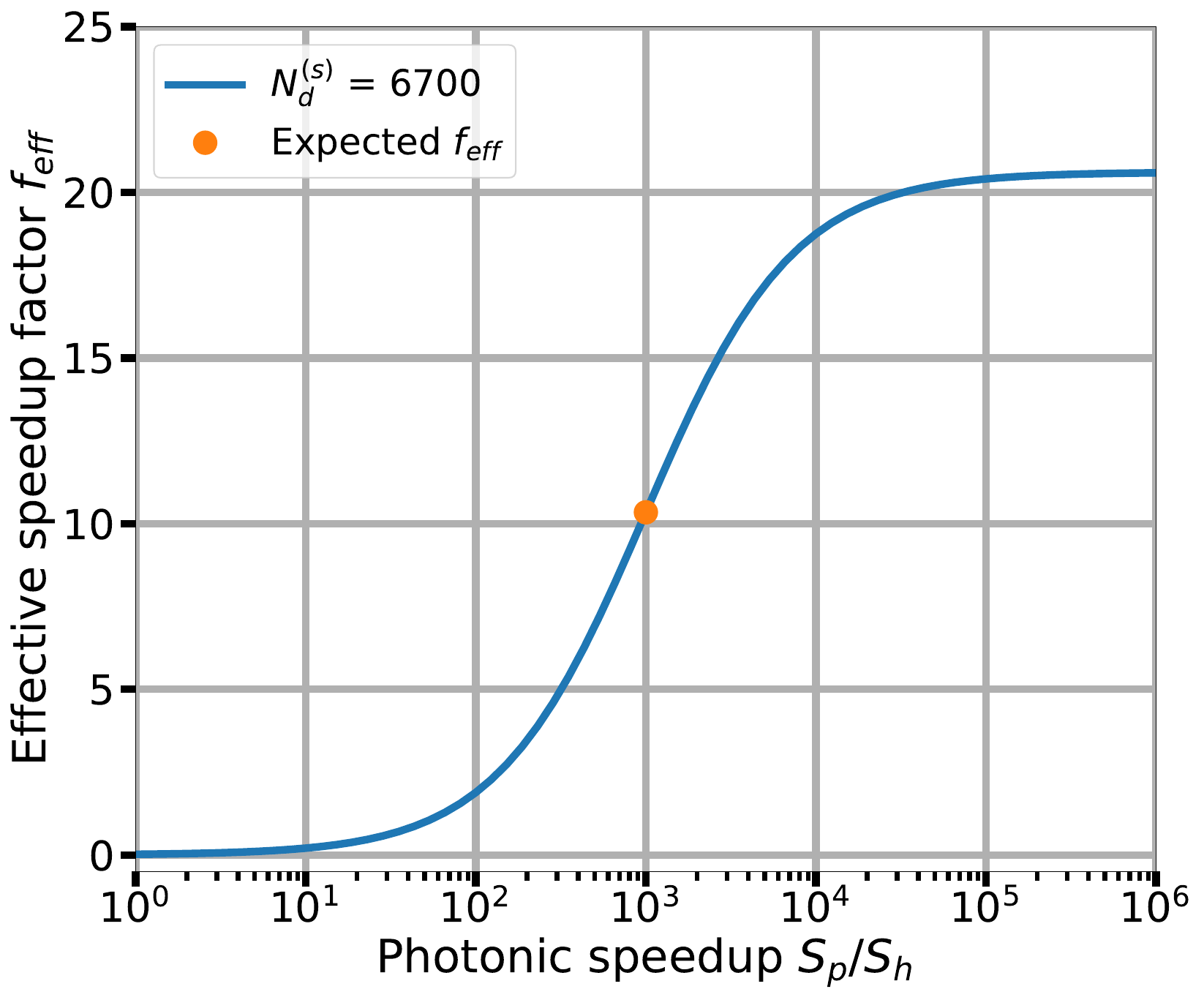}
    \caption{The effective speedup factor in \labelcref{eq:effective_speed2} is plotted as a function of the photonic speedup respective to the half precision calculations. If the photonic speedup is three orders of magnitude, then we expect an effective speedup factor of one order of magnitude.}
    \label{fig:speedup_MPCG}
\end{figure}

Therefore, we expect the effective speedup factor to increase as the matrix becomes harder to invert, which depends on the size and eigenvalues of the Dirac matrix. We use conservative hypothetical values based on \Cref{table:1} to show the potential of these photonic accelerated inverters. 
For lattice QCD in four dimensions and with the even and odd preconditioner, standard CG can require more than 6700 iterations to converge~\cite{Clark:2009wm}. Hence, we use the value $N_d^{(s)} = 6700$. 
For the calculation time with MPCG, we assume $I^{(mp)} = 1$ and $N_d^{(mp)} = 12$ and for the calculation time of the multilayered MPCG $I^{(ml)} = 32$, $N_d^{(ml)} = 12$ and $N_i^{(ml)} = 120$. 
With these values we plot the effective speedup factor in \labelcref{eq:effective_speed2} as a function of the photonic speedup respective to the half precision calculations $S_p/S_h$ in \Cref{fig:speedup_MPCG}. 
We can see that if the photonic accelerator are 1000 times quicker than the half precision calculations of the NVIDIA A100, then we get an order of magnitude of speed using photonic accelerators. 

This is only an estimation, and many technical details and optimisation need to be solved to claim a more accurate estimation. The important task is to develop methods that keep $N_d$ and $N_i$ low to extract speed from the photonic accelerated inverters. For this, we need to choose the right combination of linear solver, mixed precision method, preconditioners, and photonic accelerator.

\subsection{Intuitive interpretation of the photonic accelerated inverters}
\label{sec:IntuitiveInter}

Before we close this section, we want to give an intuitive interpretation of these methods, which could open the path to photonic accelerated inverters. This only has the purpose of helping develop an intuition, and is not exactly what happens. 
\begin{equation}
\begin{aligned}
\left|x_{(0)}\right| & =[00] .00000000 \\
& \quad \quad  \rightarrow \\
\left|x_{(9)}\right| & =29 .[16] 005376 \\
& \quad \quad \quad \   \rightarrow \\
\left|x_{(41)}\right| & =29.444[06] 173 \\
& \quad  \quad  \quad  \quad \quad \  \rightarrow \\
\left|x_{(117)}\right| & =29.44458438[\ldots]
\end{aligned}
\label{eq:sliding}
\end{equation}
We can look at the evolution of the norm of the guess $x_{(i)}$ at one of the inversions in \labelcref{eq:sliding} after certain numbers of iterations $i$. We can see that the algorithm moves quickly from the initial guess of 0 to a value around 29-30. And then the algorithm starts to take smaller and smaller steps towards the solution. Therefore, we can interpret the algorithm as sliding through the digits of the solution, where the scaling step is essential to deal with these smaller steps.  The limited dynamical range limits the space that the algorithm can explore at each run of low-precision CG. When the high-precision correction step is performed, and thanks to the scaling step, we then explore a new limited space of smaller values. The box (square brackets) represents the limited exploration range of the low-precision iterations and the sliding represents the high-precision correction step. This is iterated until the algorithm converges to the desired accuracy. The size of the box depends on the accuracy of the calculations and the eigenvalues of the Dirac matrix and the sliding speed depends on the calculation speed. 

The combination of a bad condition number in \labelcref{eq:condition_number} and low accuracy can make the box too small, making the algorithm fail. However, this can be alleviated with better mixed precision methods and by using preconditioners, as explained in \Cref{sub:condition}. This would represent an increase in the size of the box. 

Furthermore, the multilayered MPCG would be represented by a small box and a bigger box. The small box slides inside the bigger box thanks to the intermediate-precision correction steps, and once the small box reaches the edge of the bigger box, the bigger box slides thanks to the high-precision correction steps to move closer to the desired accuracy. The advantage compared to the standard MPCG is that the small box moves quicker inside the bigger box than by itself. The bigger box can be seen as the GPUs accelerating the calculations, and the small box as the photonic accelerator complementing the GPUs to achieve even quicker calculations.

\section{Conclusion and outlook}
\label{sec:Conclusion}

The photonic accelerated inverters (PAIs) presented here could help alleviate the typical bottleneck of fermionic lattice calculations, which are matrix inversions. These PAIs could overcome the limitation in the accuracy of the photonic calculations using mixed precision methods, which was shown in \Cref{fig:iterations}. There, we can see that the number of iterations needed by the algorithms increases due to the inaccuracy. Nevertheless, if the photonic technology fulfils its potential and reaches several orders of magnitude of speed compared to its digital counterpart, this could reduce the computation time by one order of magnitude. Furthermore, the potential improvement in energy-efficiency provided by photonic accelerators is an aspect, which needs to be investigated in the future.

This is only a proof of concept, and there are many factors and technical details, which will in the end determine the power of these methods. To construct optimal PAIs, we need to find the right combination of linear solver, mixed precision method, preconditioners, and photonic accelerator. For example, we could use the Richardson's method instead of the CG method as the low-precision linear solver~\cite{Zhu2023}. There is research on using this method for implementing photonic iterative processors to directly solve the matrix inversion in the optical domain \cite{chen2023ioefficient,Chen:22}, which could further unlock the potential of the photonic technology by reducing the communication between optical and electronic components.  

Additionally, there is research on finding optimal preconditioners for GPUs~\cite{tu2018solving}, and it may be interesting to explore preconditioners, which are well-suited to exploit the strengths of the photonic technology.

Hopefully, this will ignite the interest of both, the lattice and the photonic computing communities, and lead to research from which both communities benefit.

\bibliography{refs}

\end{document}